\documentstyle[sprocl]{article}
\def\Journal#1#2#3#4{{#1} {\bf #2}, #3 (#4)}

\def\PLB{{\em Phys. Lett.}  B}
\def\PRL{\em Phys. Rev. Lett.}
\def\PRD{{\em Phys. Rev.} D}
\def\be{\begin{equation}}
\def\ee{\end{equation}}

\begin{document}

\title{Update on Open Universe Inflationary Models}

\author{ J.D. Cohn}

\address{Depts of Physics and Astronomy, University of Illinois
at Urbana-Champaign, Urbana, IL 61801}

\maketitle\abstracts{An overview of some new results in open inflation
over the past year, including the calculation of gravity wave
contributions to the Cosmic Microwave Background.}

In the past year, there has been significant progress in the understanding
of the predictions and properties of open universe inflation.
For background on earlier work and an introduction to these models
see the original papers,\cite{BGT} reviews,\cite{revs}
and references therein.
Some key features of the
earlier work are the following.  Open
universe inflationary models nucleate a bubble during inflation to generate
curvature.  The amount of inflation
after nucleation of the bubble determines $\Omega$, usually put
in by hand (but there are exceptions\cite{vil-win}).
Modifications of the original model include two field models,\cite{lin2}
where one field generates inflation and one field nucleates the bubble.
The Cosmic Microwave Background (CMB) 
temperature fluctuations for scalars include both subcurvature
and supercurvature\cite{yts} scale perturbations, 
with the latter enhanced\cite{st}
when the ratio of false to true vacuum energies is large.
In the absence of gravity there was in addition a
mode corresponding to fluctuations of the bubble wall itself
(see the references in above reviews).
Parameters such as properties of the false vacuum before
tunneling and the wall profile have been varied in these models.
These variations only 
influenced CMB temperature predictions on very large angular scale,
changing only the overall normalization for open model
large scale structure predictions.\cite{lss}

Recent developments include the calculation of gravity waves (tensor
fluctuations of the background metric) for open universe models.
Earlier unpublished work by Allen and Caldwell
found an infinite CMB contribution from gravity waves originating
in the natural de Sitter invariant vacuum.  
Defining
$$
{\delta T(\hat{n}) \over T} {\delta T(\hat{n}^\prime) \over T}
= {1 \over 4 \pi} \sum_\ell (2 \ell +1) C_\ell P_\ell (\cos \theta) \; ,
$$
with $
\hat{n} \cdot \hat{n}^\prime = \cos \theta$,  
they found $C_\ell  = \infty$ for
all $\ell$, all angular scales.  
For open universes created via bubble nucleation, the
gravity waves originate in the de Sitter invariant (false) vacuum,
but travel through the bubble wall to reach us.
The energy difference across the bubble wall
regulates the divergence found by Allen and Caldwell.\cite{ts-gw,bc}
There have been efforts to regulate this divergence
without introducing a bubble wall, unsuccessful so far.\cite{robc}
Thus, given our current understanding of the vacuum, the bubble is needed
not only to provide the open universe but to make sure that the open
universe has finite gravity waves, and
any non-bubble open universe scenario
needs some other mechanism to create finite gravity waves.

A second property\cite{ts-gw} of the gravity waves
is that the lowest momentum gravity wave
contribution is degenerate with the 'would-be' wall mode,
the bubble wall fluctuations in the absence of gravity.
This combined mode seems non-normalizable.
Although not completely conclusive, a 
rigorous and detailed quantization of the coupled scalar field and
gravity system supports the absence of these modes as well.\cite{tsgm}
In the coupled gravity wave and bubble wall system
the bubble wall oscillates as the
gravity waves go across it.\cite{ii}  A relation between gravity
waves and would-be wall fluctuations is also
useful for a heuristic
description of the CMB gravity wave spectrum
(for chaotic inflation models inside the
bubble), where the tensor contribution can be written as
a ``residual'' plus a ``wall'' contribution\cite{YTS}
$$
C_\ell^{T} =
{1 \over H_T^2}\tilde{C}_\ell^{T, {\rm min}}(\Omega) + 
{A(model) \over H_T^2} 
\tilde{C}_\ell^{\rm wall}(\Omega)\;, \; H_T = {\rm Hubble \;
constant} \; .
$$

There is a minimal gravity wave contribution,\cite{hu-white} with
the gravity waves suppressed
by a factor of $V^\prime /V$ relative to the scalars.
The tensor to scalar ratio has dependence on $\Omega$ and
model parameters,\cite{juan-gw} unlike flat models.
As the gravity wave spectrum can be 
steep at low $\ell$, and
large, requiring agreement with observations constrains models.
Parameter ranges for theories satisfying the gravity wave constraints
were given for chaotic inflation\cite{YTS} and for\cite{juan-gw} recently
proposed models of tilted open hybrid inflation\cite{gl} and 
dilaton/scalar\cite{green} inflation.  Open model scalar
and tensor contributions can now be calculated with
the publicly available
Boltzmann code CMBFAST\cite{uros,hswz} (which includes the four year
COBE normalization\cite{bw}).  The CMB polarization 
gives additional information and
has been calculated for some open models.\cite{hswz}

Another area of study was the properties of
the open universe backgrounds.  In the simplest two field models,\cite{lin2}
inhomogeneities can arise from the evolution of the second field
while the first field is driving inflation.\cite{gm} 
In addition,\cite{mgg}
the supercurvature modes in several two field models
destroy the background homogeneity, restricting the
second stage of inflation to finite regions inside the bubble.
Calculating the observationally viable range of parameters 
for these quasi-open models is in progress.\cite{mgg}
Remaining questions for these models include the likelihood of
motivating the appropriate field theory potentials from a high
energy physics theory such as string theory.

I thank A. Liddle and M. White 
for conversations and the organizers for a stimulating and
exciting conference.  This work was supported by an ONR grant and
NSF-PHY9722787.

\end{document}